\documentclass{sig-alternate-05-2015}
\usepackage{color}
\usepackage{mathtools}
\usepackage{algorithm}
\usepackage{tabularx}
\usepackage{algpseudocode}
\usepackage{amsmath} %maths

\newcommand{\pushright}[1]{\ifmeasuring@#1\else\omit\hfill$\displaystyle#1$\fi\ignorespaces}
\renewcommand{\algorithmicrequire}{\textbf{Input:}}
\renewcommand{\algorithmicensure}{\textbf{Output:}}
\newcommand\myeq{\stackrel{\mathclap{\normalfont\mbox{def}}}{=}}

 \usepackage{natbib}
 \renewcommand{\cite}{\citep}

\providecommand{\inv}{^{-1}}
\providecommand{\Abar}{\bar{\mA}}
\DeclareMathOperator*{\argmin}{argmin}

\newcommand{\lem}{\textsc{Lemon}\xspace}

\newcommand{\eps}{\varepsilon}

\newcommand{\vpara}[1]{\vspace{0.1in}\noindent\textbf{#1 }}

\usepackage[utf8]{inputenc}

\usepackage{amsfonts,amsmath,amssymb}

\usepackage{array}

\graphicspath{{./figures/}}

\DeclareMathOperator{\cut}{cut}
\DeclareMathOperator{\vol}{vol}

\providecommand{\mA}{\ensuremath{\textbf{A}}}

\providecommand{\mD}{\ensuremath{\textbf{D}}}

\providecommand{\mI}{\ensuremath{\textbf{I}}}

\providecommand{\mL}{\ensuremath{\textbf{L}}}

\providecommand{\mP}{\ensuremath{\textbf{P}}}

\providecommand{\mV}{\ensuremath{\textbf{V}}}

\providecommand{\ve}{\ensuremath{\textbf{e}}}
\providecommand{\vf}{\ensuremath{\textbf{f}}}

\providecommand{\vp}{\ensuremath{\textbf{p}}}

\providecommand{\vs}{\ensuremath{\textbf{s}}}

\providecommand{\vx}{\ensuremath{\textbf{x}}}
\providecommand{\vy}{\ensuremath{\textbf{y}}}
\providecommand{\vz}{\ensuremath{\textbf{z}}}

\RequirePackage{graphicx}
\RequirePackage{tabularx}
\RequirePackage{booktabs}
\RequirePackage{ragged2e}
\RequirePackage{xcolor}
\RequirePackage{textcomp}
\RequirePackage{xspace}
\RequirePackage{microtype}
\RequirePackage{listings}

\colorlet{TufteRed}{red!80!black}
\definecolor{halfgray}{gray}{0.55}

\definecolor{subtleblue}     {rgb}{0.02,0.04,0.48}
\definecolor{subtlered}      {rgb}{0.65,0.04,0.07} % RGB 165,10,18
\definecolor{subtlegreen}    {rgb}{0.06,0.44,0.08}
\definecolor{subtledarkblue} {rgb}{0,.1,.6}
\definecolor{lightsubtleblue}{rgb}{0,.4,.6}
\definecolor{ecru}           {rgb}{1.0,.98823,.95686}

\definecolor{stanfordred}      {rgb}{0.6431,0.000,0.1137}

\definecolor{stanfordsandstone}{rgb}{0.9059,0.8196,0.6039}
\definecolor{stanfordblue}     {rgb}{0.1451,0.5176,0.7333}
\definecolor{stanfordgreen}    {rgb}{0.1608,0.3961,0.2863} % rgb 41/101/73
\definecolor{stanforddarkgray} {rgb}{0.2627,0.2902,0.2667} % rgb 67/74/68
\definecolor{stanfordlightgray}{rgb}{0.8392,0.8667,0.8275} % rgb 214/221/211
\definecolor{stanforddarkgreen}{rgb}{0.2353,0.2118,0.1373}
\definecolor{stanforddeepred}  {rgb}{0.6510,0.2275,0.0000}
\definecolor{stanfordneutralkhaki}{rgb}{0.5686,    0.5333,    0.4510}
\definecolor{stanfordbrightgreen}{rgb}{0.0039 ,   0.5137  ,  0.3725}
\definecolor{stanfordbrightblue}{rgb}{0.1451,0.5176, 0.7333}
\definecolor{stanfordbrightseagreen}{rgb}{0.0000,0.5020,0.5529} % rgb 0/128/141
\definecolor{stanfordbrightyellow}{rgb}{0.9412,0.6863,0.0000} % rgb 240/175/0
\definecolor{stanfordbrightwine}{rgb}{0.2353,0.0667,0.0275}

\newcommand{\PreserveBackslash}[1]{\let\temp=\\#1\let\\=\temp}
\newcolumntype{L}[1]{>{\PreserveBackslash\RaggedRight}m{#1}}
\newcolumntype{M}[1]{>{\PreserveBackslash\RaggedRight}p{#1}}
\newcolumntype{R}[1]{>{\PreserveBackslash\RaggedLeft}m{#1}}
\newcolumntype{S}[1]{>{\PreserveBackslash\RaggedLeft}p{#1}}
\newcolumntype{Z}[1]{>{\PreserveBackslash\Centering}m{#1}}
\newcolumntype{A}[1]{>{\PreserveBackslash\Centering}p{#1}}

\newcolumntype{U}{>{\setlength{\RaggedRightParindent}{0pt}\RaggedRight\arraybackslash\noindent}X}
\newcolumntype{V}{>{\RaggedLeft\arraybackslash}X}
\newcolumntype{W}{>{\Centering\arraybackslash}X}

\graphicspath{{.}{./figures/}}

\lstnewenvironment{pseudocode}[1][]{%
  \lstset{language=python,#1%             % use our version of highlighting
  }%
}{}

\lstdefinelanguage{matlabfloz}{%
  alsoletter={...},%
  morekeywords={%                             % keywords
  break,case,catch,continue,elseif,else,end,for,function,global,%
  if,otherwise,persistent,return,switch,try,while,...,ones,zeros,eye},%
  comment=[l]\%,%                             % comments
  morecomment=[l]...,%                        % comments
  morestring=[m]',%                           % strings
}[keywords,comments,strings]%

\lstnewenvironment{dispmcode}{%
    \lstset{language=matlabfloz,%
      numbers=none,frame=none,%
      keywordstyle=\color{subtleblue},%,          % keywords
      commentstyle=\color{thegreen},%         % comments
      stringstyle=\color{subtleed}%             % strings
    }%
  }%
  {}

\lstnewenvironment{dispmresults}{%
    \lstset{language=matlabfloz,%
      numbers=none,frame=none,backgroundcolor=\color{ecru},%
      keywordstyle=\color{theblue},%,          % keywords
      commentstyle=\color{subtlegreen},%         % comments
      stringstyle=\color{subtlered}%             % strings
    }%
  }%
  {}

\lstnewenvironment{numberedmcode}{%
    \lstset{language=matlabfloz,%
      numbers=left,frame=none,%
      keywordstyle=\color{subtleblue},%,          % keywords
      commentstyle=\color{subtlegreen},%         % comments
      stringstyle=\color{subtlered}%             % strings
    }%
  }%
  {}

\begin{document}

% Copyright
%\setcopyright{acmcopyright}
%\setcopyright{acmlicensed}
%\setcopyright{rightsretained}
%\setcopyright{usgov}
%\setcopyright{usgovmixed}
%\setcopyright{cagov}
%\setcopyright{cagovmixed}

%Conference
\conferenceinfo{}{}
%\acmPrice{\$15.00}

% DOI
%\doi{10.475/123_4}
\doi{N/A}

% ISBN
%\isbn{123-4567-24-567/08/06}
\isbn{N/A}

%
% --- Author Metadata here ---
%\conferenceinfo{KDD}{'16 San Francisco, California USA}
%\CopyrightYear{2007} % Allows default copyright year (20XX) to be over-ridden - IF NEED BE.
%\crdata{0-12345-67-8/90/01}  % Allows default copyright data (0-89791-88-6/97/05) to be over-ridden - IF NEED BE.
% --- End of Author Metadata ---

\title{ Scalable and Robust Local Community Detection\\ via Adaptive Subgraph Extraction and Diffusions
% Subgraph Extraction and Seed Set Augmentation for \\Robustifying Local Network Community Detection
%\titlenote{(Produces the permission block, and copyright information). For use with SIG-ALTERNATE.CLS. Supported by ACM.}
}

\numberofauthors{2}
\author{
% 1st. author
\alignauthor
Kyle Kloster%\titlenote{Foot note here}
\\
       \affaddr{NCSU}\\
       \email{kakloste@ncsu.edu}
% 2nd. author
\alignauthor
Yixuan Li
\\
       \affaddr{Cornell University}\\
       \email{yli@cs.cornell.edu}
}

\maketitle

\begin{abstract}
Local community detection, the problem of identifying a set of relevant nodes nearby a small set of input seed nodes, is an important graph primitive with a wealth of applications and research activity.
Recent approaches include using local spectral information, graph diffusions, and random walks to determine a community from input seeds.
As networks grow to billions of nodes and exhibit diverse structures, it is important that community detection algorithms are not only efficient, but also robust to different structural features.

Toward this goal, we explore pre-processing techniques and modifications to existing local methods aimed at improving the scalability and robustness of algorithms related to community detection. Experiments show that our modifications improve both speed and quality of existing methods for locating ground truth communities, and are more robust across graphs and communities of varying sizes, densities, and diameters.
Our subgraph extraction method uses adaptively selected PageRank parameters to improve on the recall and runtime of a walk-based pre-processing technique of Li et al.~\cite{li2015localspectral} for extracting subgraphs before searching for a community.
We then use this technique to enable the first scalable implementation of the recent Local Feidler method of Mahoney et al.~\cite{mahoney2012localspectral}. Our experimental evaluation shows our pre-processed version of Local Fiedler, as well as our novel simplification of the LEMON community detection framework of Li et al.~\cite{li2015localspectral}, offer significant speedups over their predecessors and obtain cluster quality competitive with the state of the art.

% We focus on two related goals: a subgraph extraction technique aimed at obtaining a medium-sized, high-recall region near the seeds to enable efficient use of more sophisticated methods on the subgraph that would be intractable on the full network; and a seed augmentation procedure aimed at identifying a few nodes with high precision near the seeds, to obtain a more robust seed set to improve other local network analyis tasks.
% We evaluate our methods alongside related, recently proposed pre-processing techniques and demonstrate significant improvement for both tasks.
% Comprehensive experiments reveal close connections of algorithmic performance to the diameter and edge density of communities; in particular, while it is widely known that social networks have small diameter, we observe that communities therein have extremely small diameter, which improves the performance of subgraph extraction and justifies why local expansion approaches are well-suited for the task of community identification.
% Finally, we gain insight into a successful recent diffusion method for local community detection that
% inspired our proposed techniques, and demonstrate that our simplified, more efficient algorithm can produce clusters of comparable quality.

\end{abstract}

\begin{CCSXML}
<ccs2012>
<concept>
<concept_id>10002950.10003624.10003633.10010917</concept_id>
<concept_desc>Mathematics of computing~Graph algorithms</concept_desc>
<concept_significance>500</concept_significance>
</concept>
</ccs2012>

\end{CCSXML}

\ccsdesc[500]{Mathematics of computing~Graph algorithms}


%  Use this command to print the description
\printccsdesc

\keywords{community detection; semi-supervised learning; graph algorithms; local algorithms}

\section{Introduction}\label{sec:introduction}

We consider the general problem of identifying a community around a given seed node or nodes of interest. That is, given an input node (or nodes) in the graph, we conider the goal of trying to find a cluster of which the node is a member.

An evolving research direction in graph data mining is to develop community detection algorithms that scale to extremely large graphs, e.g., algorithms that rely on local computations involving {\em only} nodes relatively close to the given seeds. For example, recent work on \emph{local} graph diffusion methods have shown promise that one can find clusters in a localized way, i.e. without looking at most of the graph. The general framework operates by first computing a diffusion vector, then returning as the detected community the set of nodes that have largest mass. These diffusion procedures can be viewed as propagating large probability values from the labeled nodes (``{\em{seeds}"}) to the remaining unlabeled nodes, which is the key ingredient in many local graph diffusion algorithms~\cite{andersenlang2006communities,andersen2006local,avron2015community,gleich2015using,kloster2014heatkernel,li2015overlapping,li2015localspectral,spielman2008local}.
A useful local diffusion process is one that effectively propagates probabilities to the nodes that are most relevant to the given seeds, without mixing to the entire graph.

%Several graph diffusion mechanisms have been studied for the purpose of community detection, including personalized PageRank~\cite{andersen2006local}, personalized heat kernel~\cite{chung2009local,kloster2014heatkernel}, time-dependent PageRank~\cite{avron2015community,Gleich-2014-dynamic-pagerank} and short-length random walks~\cite{andersenlang2006communities}.

Though these approaches have achieved some success, cluster quality and runtime can depend heavily on features of a given dataset like density and community diameter. For example, a simple method like a short random walk might be effective on a dataset with small diameter communities, but could fail to reach a large portion of a community in a sparser, larger diameter graph. On the other hand, a more sophisticated method like a local spectral approach could maintain consistent cluster quality across these graphs, but slow down orders of magnitude as the size of the graph increases.

To combat this, we present a pre-processing technique that is robust in the face of such varied characteristics. The technique uses PageRank with adaptive parameters to consistently extract a high recall subgraph of medium size. Comprehensive experiments show that our PageRank-based technique attains the best recall among algorithms considered, especially on networks in which communities have larger diameter. This improves over the $k$-walk approach used in~\cite{li2015localspectral}, which fails to expand to much of the larger-diameter communities.

The subgraph extraction method enables efficient use of more sophisticated methods on the subgraph without losing large portions of valuable graph regions.
 Our hope is this technique will combine with other community detection and other semi-supervised learning algorithms. As a simple case study, we show our technique enables the first efficient use of the recently proposed locally-biased Fiedler vector method (MOV)~\cite{mahoney2012localspectral} on large datasets such as social networks from the SNAP repository~\cite{snapnets}.

 %We extensively evaluate various diffusion methods including short-step random walk (used in~\cite{andersen2006local, li2015localspectral}) and propose alternative subgraph extraction methods using personalized PageRank and heat kernel diffusion.
 Finally, we make modifications to the local spectral method of Li et al.~\cite{li2015localspectral} to produce a much simpler and faster method that still attains competitive cluster quality.
 We demonstrate significant improvement on 11 different real-world network spanning various domains of network application (see Table~\ref{tab:data-summary}).

% \vpara{Seed set augmentation with good precision}
% For many semi-supervised algorithms to work well, they require a sufficiently large set of seeds as input. However, in practice obtaining seed sets larger than a single node can be  expensive or, in some cases, impossible. In many other cases, especially when no pre-defined community exists for a node of interest, augmenting the initial seed set is desirable to provide addition prior knowledge to the detection algorithm. Therefore, augmenting the initial seed set can be widely useful in any graph mining algorithms whose performance improves with larger seed sets.
%
% We address this challenge common to many local community detection methods: how can we use a single seed node to a small seed set with {\em high precision}? Our experiments demonstrate that, using the top ranked nodes produced by a diffusion method (e.g. personalized PageRank) on a whole graph often gives poor precision in growing a single seed into a small seed set. However, we find that using a subgraph extraction method as a pre-processing routine before ranking nodes near the seed yields much better precision in identifying new ground truth nodes. Our method can attain an average precision of \TODO{fill in the number} when growing a seed into a set of three nodes, for example.

\vpara{We summarize our main contributions as follow:}
\begin{enumerate}
\item We propose and systematically evaluate different subgraph extraction techniques for improving the performance of any local graph analysis algorithms. We show our proposed modification improves recall.
\item We make substantial simplifications to the framework of Li et al. to yield a simpler, yet much faster implementation.
\item We present experiments on 11 datasets spanning various domains of network applications and observe how the performance of subgraph extraction algorithms relate to certain community and graph properties like diameter and edge-density, leading to a better understanding of what kind of algorithm is best to employ on different categories of networks.
\item We investigate the novel problem of seed set augmentation and evaluate common community detection tools for obtaining a small, high-precision set of seed nodes.
\end{enumerate}

We make our experimental codes available in the spirit of reproducible research: \url{https://github.com/kkloste/lemon-sqz}.

\section{Related Work}\label{sec:related-works}

Much recent work has studied network and community properties and algorithmic approaches related to finding communities from seed sets.
One study identified common structural properties of ground truth communities, like separatedness and internal cohesion~\cite{yang2012definingevaluating}, while another observed that characteristics such as conductance and diameter can be seen as classification features that can distinguish between ground truth communities and the outputs of a variety of community detection algorithms~\cite{abrahao2012separability}. Kloumann and Kleinberg~\cite{kloumannkleinberg2014community} studied the impact of seed set characteristics on the quality of clusters produced by personalized PageRank diffusions.
While it is widely known that social networks exhibit small and even shrinking diameter~\cite{leskovec2005graphs}, it has also been found that spectral methods produce clusters that have small diameter~\cite{leskovec2009communitystructure}. A related study found that egonets (node neighborhoods, which by definition have a diameter of at most 2) can be good clusters, in the sense that they can have low conductance~\cite{gleichseshadhri2012vertex}.

%These results have shown that these properties -- compactness, small diameter, low conductance, high internal edge cohesion -- all seem desirable and sometimes common characteristics in clusters.
From the algorithmic side, a swath of new methods have appeared for identifying local communities without having to look at the entire network structure.  Methods for \emph{locating} local network community structure from a given seed set have included diffusions, Monte Carlo methods, and spectral approaches.
Recent diffusion vectors include random-walk vectors~\cite{andersenlang2006communities, spielman2008local}, the personalized PageRank vector~\cite{andersen2006local,jeh2003scaling-personalized} heat kernel vector~\cite{chung2007heat,kloster2014heatkernel}, and the time-dependent PageRank vector~\cite{avron2015community,Gleich-2014-dynamic-pagerank}. (We briefly review the vectors most relevant to our study in Section~\ref{sec:background}.)
Local spectral-based approaches include a locally-biased version of the Fiedler vector~\cite{mahoney2012localspectral}, as well as the recent \textsc{Lemon} vector~\cite{he2016local,li2015localspectral}.

\section{Preliminaries}\label{sec:prelims}

Here we fix our notation and review recent approaches related to local community detection. Let $G = (V,E)$ be a graph with $n = |V|$ nodes and $m = |E|$ edges. We assume that the graph is unweighted and undirected, which is commonly assumed in the context of community detection. We denote the adjacency matrix associated with the graph $G$ by $\mA$, with entries $a_{ij}=1$ if nodes $i$ and $j$ are connected and $a_{ij}=0$ otherwise. Let $\mD$ be the diagonal matrix of node degrees where $\mD_{ii} = d(v_i)$, and $\mP = (\mD^{-1}\mA)^{T} = \mA\mD^{-1}$ be the random walk transition matrix, and note that in our notation it is \emph{column} stochastic.
Lastly, $\mL = \mD - \mA$ is the combinatorial Laplacian.

We assume vectors are \emph{column} vectors. For a fixed node $j$ we denote by $\ve_j$ a standard basis vector  with a 1 in the $j$th entry, and $\ve$ denotes the vector of all 1s.
We consider applications that seek information about a small set of input nodes which we call ``seed" nodes. For a node set $S$, $\ve_S$ denotes the indicator vector of the set $S$,  i.e. $\ve_S$ is all 0s except with 1s in entries corresponding to nodes in $S$.

%Many data mining tasks interested in learning more about seed nodes attempt to discover a ``community" of nodes nearby the seed set.
%Such a group of nodes can be used in label propagation, link prediction, community detection, and other semi-supervised learning tasks.
Proposed definitions and properties of communities vary widely, though an often-given intuition for a community is ``a set of nodes with high internal connectivity, and relatively lower external connectivity". A related and commonly-adopted metric for evaluating how much a node set ``resembles" a community is {\em conductance}, which is defined as follows. Given a set of nodes $C\subseteq V$, the conductance of $C$ is
\begin{equation}
\phi(C) = \frac{\cut(C,\bar C)}{\min\{\vol(C), \vol(\bar C)\} },
\end{equation}
where $\bar C=V \backslash C$ consists of all nodes not in $C$, $\cut(X,Y)$ denotes the number of edges between the nodesets $X$ and $Y$, and $\vol(X) = \sum_{v\in X} d(v)$, i.e. the ``edge volume" of the nodeset $X$. Conceptually, $\phi(C)$ is the probability that a random walk of length one will escape $C$, given that we start from an edge-endpoint chosen uniformly at random inside $C$.

%Because conductance has been widely used to measure how ``community-like" a cluster is, many algorithms for community detection output a set of nodes by optimizing conductance. All of the algorithms which we treat in this paper involve conductance to some extent. One other concept relevant to many tools for community discovery, and in particular all of the algorithms we consider in this paper, is that of a graph diffusion. Next we describe this concept.

\vpara{Diffusion vectors.} A graph diffusion is a probability vector of the form
\begin{equation}
\vf = \sum\limits_{k=0}^{\infty} c_k\mP^k\vp_0,
\end{equation}\label{eqn:general_diffusion}
where the coefficients $c_k$ are any values such that $c_k \geq 0$, $\sum_{k=0}^{\infty} c_k = 1$, and $\vp_0$ is the initial distribution of probability across the nodes defined by
$$\vp_0 =
\begin{cases}
    d(v) / \sum_{v\in S} d(v)      & \quad \text{if } v \in S\\
    0  & \quad \text{otherwise}.\\
  \end{cases}
$$
Some approaches also use the weighting $\vp_0 = \ve_S / |S|$ in place of the above degree-weighted normalization. We call the terms $c_k$ the \emph{diffusion coefficients}.  A particular choice of diffusion coefficients is essentially an assignment of weights to random walks of different lengths, and hence a way to emphasize nodes at specific walk depths. %In this work we are concerned primarily with \emph{local} diffusions, i.e. for which $\vp_0 = \ve_S /|S| $ for a seed set $S$.

\vpara{Sweep procedure.}
Once the estimation of a diffusion from a seed set $S$ is obtained, one can produce a small conductance community via the so-called {\em sweep} procedure. A sweep over a vector involves sorting the nodes in descending order according to the entries in the vector, and computing the conductance of each prefix of the sorted list. The set found to have smallest conductance is then returned by the sweep procedure as the detected community for the given seed set. %In the case of a tie, the set with minimal conductance and fewest nodes is returned.

\section{Local Diffusion Methods}\label{sec:background}

In this section we describe several recent methods for local community detection, which we select for their widespread attention in the literature and because of the diversity in characteristics of the clusters that they identify. %For a description of the characteristics of the clusters output by these algorithms, see Section~\ref{sec:char}.
We give a brief overview of the local graph diffusions including random walks, personalized PageRank (PPR)~\cite{andersen2006-local}, heat kernel (HK)~\cite{chung2009local}, and local spectral methods MOV~\cite{mahoney2012localspectral} and LEMON~\cite{li2015localspectral}.

\subsection{Overview of algorithms considered}\label{sec:diffusion-overview}

%Below we briefly summarize local diffusion-based approaches, including $k$-walk diffusion vectors, the heat kernel diffusion (HK)~\cite{chung2009local}, and personalized PageRank (PPR) vector~\cite{andersen2006-local}. They are each special instances of the general graph diffusion presented in Equation~\ref{eqn:general_diffusion}.

%\vpara{Walk based diffusion vectors.}
\vpara{The $k$-walk diffusion vector.}
The $k$-walk diffusion vector is a graph diffusion in which a single diffusion coefficient in the formulation~\eqref{eqn:general_diffusion} is 1, i.e., $c_k=1$, and $c_j=0$ for all other $j \neq k$. In other words, $\vf = \mP^k\vp_0$ for some $k \in \mathbb{N}$.
%The vector is defined by
%\begin{equation}
%RW_k = \mP^k\vp_0,
%\end{equation}
%where $\vp_0$ is the initial probability distribution vector (i.e. some stochastic vector). In the context of tasks related to local community detection, the diffusion starts at a small seed set $S$, and
%$$\vp_0 =
%\begin{cases}
%    d(v) / \sum_{v\in S} d(v)      & \quad \text{if } v \in S\\
%    0  & \quad \text{otherwise}.\\
%  \end{cases}
%$$
%Such random walk vectors have been shown to obtain sets of good conductance~\cite{andersenlang2006communities}.
%Some approaches also use the weighting $\vp_0 = \ve_S / |S|$ in place of the above degree-weighted normalization. %In the common case that $S$ is a single node, the two normalizations are equivalent.
%

\vpara{The personalized PageRank diffusion vector.}
For a fixed $\alpha \in (0,1)$, the personalized PageRank vector can be defined as
\begin{equation}
PPR =   (1-\alpha) \sum\limits_{k=0}^{\infty} \alpha^k \mP^k\vp_0.
\end{equation}\label{eqn:ppr}
The personalized PageRank vector can be interpreted as the stationary distribution of a random walk with restart.
The diffusion was proposed in~\cite{jeh2003scaling}, but we use the algorithm presented in~\cite{andersen2006local}.
%The properties of PPR have been extensively studied and widely applied in ranking algorithms.
%Andersen et al.~\cite{andersenlang2006communities} described an efficient local algorithm for computing the PPR vector. We use an implementation of that algorithm provided by~\cite{gleich2012vertex}.
%The properties of this diffusion vector have been extensively studied and widely applied in ranking algorithms. \cite{kloumannkleinberg2014community} used the idea of personalized PageRank diffusion and other variants of PageRank vectors for community discovery. However, they did not consider the pre-processing of subgraph sampling. We expand the work by evaluating PPR diffusion vector under a more comprehensive sets of experiments.

\vpara{The heat kernel diffusion vector.}
The heat kernel diffusion replaces the weights $c_k = (1-\alpha)\alpha^k$ with $e^{-t} (t^k / k!)$:
\begin{equation}
HK = e^{-t} \bigg(\sum\limits_{k=0}^{\infty}\frac{t^k}{k!} \mP^k \bigg) \vp_0.
\end{equation}
In contrast with PPR, the heat kernel diffusion models the spread of heat across a graph starting from a seed set.
The diffusion was proposed and analyzed in~\cite{chung2007heat}, but we use the algorithm presented in~\cite{kloster2014heatkernel}.
%First studied theoretically in~\cite{chung2007heat,chung2009local}, we use the algorithm presented in the study~\cite{kloster2014heatkernel}.

Both the PPR and HK algorithms that we study in this paper were designed to be \emph{local}: even though the diffusion vectors themselves are global (i.e. totally dense on a connected graph), the algorithms used here are approximations that explore only a small portion of the graph.
Both approximation algorithms have theoretical guarantees that limit the amount of work they perform before converging (and, hence, limit the size of the output cluster) by a small constant that depends only on the accuracy parameter $\eps$ used. These theoretical guarantees rely crucially on the fact that the approximation schemes both measure error in the degree-weighted infinity norm $\| \mD\inv(\cdot) \|_{\infty}$.

\vpara{The MOV (locally-biased Fiedler) vector.}
The MOV vector, first introduced in~\cite{mahoney2012localspectral}, offers a way to bias the standard global Fiedler vector so that it identifies a good conductance cut localized near an input set of nodes. Formally, the MOV vector is the solution $\vx$ to
\begin{align}
 \argmin\limits_{\vx}~~& \vx^T \mL \vx &\label{eqn:mov-objective} \\
 ~~~~~~~~     \text{s.t.} ~~       \vx^T\mD\vx &= 1 &\label{eqn:mov-constraint1} \\
 ~~~~~~~~                           \vx^T\mD\ve &= 0 &\label{eqn:mov-constraint2} \\
 ~~~~~~~~     (\vx^T\mD\vs)^2 & \geq \kappa, & \label{eqn:mov-constraint3}
\end{align}
where $\kappa \in (0,1)$ is an input parameter  that controls the extent to which the solution $\vx$ is localized onto the input seed set $S$ represented by the specially constructed seed vector $\vs$. In contrast with the local algorithms for the above diffusions, the seed vector here is not a sparse indicator vector $\ve_S$ but instead the following dense vector
\[
\vs = \sqrt{ \tfrac{\vol(S) \vol(\bar S)}{\vol(G)} }
\left( \tfrac{1}{\vol(S)} \ve_S  -  \tfrac{1}{  \vol(\bar S)} \ve_{\bar S} \right).
\]
Once the MOV vector $\vx$ is computed, a sweep over $\vx$ obtains a good conductance cut near the seeds $S$.
%Note that if we exclude Equation~\eqref{eqn:mov-constraint3} from the optimization problem, the solution $\vx$ to the resulting optimization problem is exactly the standard Fiedler vector; the authors introduced Equation~\eqref{eqn:mov-constraint3} as a way to force the solution $\vx$ to be biased toward the target set $S$. Interestingly, the authors prove that the solution $\vx$ is equivalent to a personalized PageRank vector seeded on $S$, but with a particular value of $\alpha$ that can't be determined a priori.
Although the MOV vector yields a localized set, the method itself is global in that computing $\vx$ requires reading the whole graph due to the density of the vector $\vs$ and the global nature of the objective function~\eqref{eqn:mov-objective}. This prevents the MOV algorithm from scaling to larger datasets; we address this concern later by using a subgraph extraction technique in pre-processing and running MOV on a much smaller subgraph (Section~\ref{sec:experiments}).

\vpara{The LEMON (local spectral subspace) method.}
%
%We also briefly review here the locally-biased spectral methods for community identification. Both the local spectral clustering method from Li et al. \cite{li2015localspectral} and Mahoney et al. \
The \textsc{LEMON} method~\cite{li2015localspectral} iteratively grows a community from input seeds using a diffusion that is computed over only a small subspace of walk vectors, such that the diffusion maximizes overlap with the iterative seed set. This subspace has been referred to as a {\em local spectral subspace}, which is simply a partial Krylov subspace. The local spectral subspace is formed from the input seed vector $\ve_S$ and a specially normalized form of the adjacency matrix defined as
\begin{equation}\label{eqn:normalized_matrix}
\Abar ~~\myeq ~~(\mD+\mI)^{-1/2}(\mA+\mI)(\mD+\mI)^{-1/2}.
\end{equation}

Given an input set $S$ and input parameters $k$ and $l$, form the local spectral subspace matrix
\begin{equation}
  \mV_{k,l} = [ \Abar^k \ve_S, ... , \Abar^{k+l-1}\ve_S ].
\end{equation}
Then the \textsc{Lemon} algorithm solves for the following $\ell_1$-norm optimization problem:
\begin{align}
\min\limits_{\vy} &~~~||\mV_{k,l}\vy||_1 \label{eqn:lemon-opt}\\
\text{s.t.} &~~~\mV_{k,l} \vy \geq 0,\\
&~~~(\mV_{k,l}\vy)(S) \geq 1,
\end{align}
where the objective function  is a regularized term with sparsity penalty. Once a solution $\vy$ is found, then $\vx = \mV_{k,l}\vy$ is the \textsc{Lemon} diffusion vector we are looking for.
%The first constraint guarantees that the resulting \textsc{Lemon} vector is nonnegative; the second constraint guarantees the seed nodes are in the support of the output vector.
%Intuitively, this optimization formulation seeks a small number of nodes that are most closely related to the current seed set.

% To interpret the optimization formulation from a geometric perspective, \textsc{Lemon} is essentially seeking a sparse diffusion vector in the span of the local spectral subspace, such that the seed is in its support. In other words, the learning objective here is a locally-biased spectral program which forces the solution to be well-connected with the known seed set $S$.

%Once $\vx$ is computed, its largest entries are added to the current seed set $S_{j}$ to form a new seed set $S_{j+1}$; then a new \textsc{Lemon} vector is computed from $S_{j+1}$. The original paper explores a number of stopping criterion; we use their ``auto" stopping mechanism, which functions as follows. After each iterative \textsc{Lemon} vector is computed, perform a sweep over $\vx$ and record the best conductance found. Once this sweep process reaches a local minimum, stop the iterative procedure, and output the community determined by the sweep procedure on the current \textsc{Lemon} vector.

%%%
%%%  DATASETS
%%%

%\section{Characteristics of datasets vs algorithmic clusters}\label{sec:characteristics}
\section{Datasets}\label{sec:datasets}

%%Having described the algorithms themselves,
%Here we discuss the qualitative differences among the outputs of the different algorithms and describe the datasets which we later use to provide quantitative differences.
%% We end the section by discussing how we intend to improve on the set of algorithms we considered in the previous section.
%We use the datasets to evaluate algorithms' abilities to uncover ground truth nodes, but also to better understand how certain features affect the performance of these algorithms. In particular, we look at the size, edge density, and diameter to see how these characteristics affect performance of diffusion-based and walk-based mechanisms.

%\subsection{Variety of dataset properties}

%Here we detail the properties of the networks and ground truth community information that we use.
%In our collection of datasets both the graphs and communities have a variety of size scales, diameters, and densities. Moreover, these networks come from variety of types % (citation, co-purchasing, social networks)
%and we perform experiments in an exhaustive manner (for example, by considering performance over all communities, and seeding on all community members) to try to gain a fuller understanding of how different algorithms behave in the presence of different network categories and characteristics. We summarize the dataset properties in Table~\ref{tab:dataset-chars} and discuss specific features in the subsections below.

\begin{table*}\label{tab:data-summary}
 \begin{centering}
\begin{tabularx}{\linewidth}{XXXXXXXXX}  \toprule
 data & abbrv. & $|V|$  & $|E|$  & number  & $\overline{|C|}$  & $\overline{d_C}$  & $\overline{d_C/d}$  & $\overline{\text{diameter}}$ \\  \midrule
 \texttt{citeseer} & \texttt{cite} & 2,110 & 3,668 & 7 & 207 & 2.9 & 0.85 & 14.3  \\
 \texttt{cora} & \texttt{cora} &2,485 & 5,069 & 8 & 273 & 3.7 & 0.88 & 11.8  \\
 \texttt{senate} &\texttt{sen} & 1,884 & 16,662 & 110 & 82 & 12.1 & 0.58 & 4.4  \\
 \texttt{usps-3nn} & \texttt{us3} &9,298 & 21,256 & 10 & 918 & 4.4 & 0.96 & 16.8  \\
 \texttt{usps-10nn} & \texttt{us10} &9,298 & 68,381 & 10 & 925 & 13.7 & 0.92 & 9.5 \\
 \texttt{amazon} &\texttt{amaz} & 334,863 & 925,872 & 2,110 & 25 & 6.3 & 0.96 & 3.9 \\
 \texttt{dblp} &\texttt{dblp} & 317,080 & 1,049,866 & 1684 & 53 & 6.0 & 0.74 & 3.5 \\
 \texttt{friendster} & \texttt{fri} &65,608,366 & 1,806,067,135 & 3,704 & 61 & 21.0 & 0.23 & 2.9 \\
 \texttt{livejournal} & \texttt{lj} &3,997,962 & 34,681,189 & 3,442 & 38 & 22.7 & 0.68 & 2.5 \\
 \texttt{orkut} & \texttt{ork} &3,072,441 & 117,185,083 & 4,571 & 236 & 29.9 & 0.31 & 4.7 \\
 \texttt{youtube} & \texttt{yout} &1,134,890 & 2,987,624 & 1,266 & 47 & 3.4 & 0.24 & 4.4 \\
% IN SUBMITTED VERSION:
% \texttt{citeseer} & \texttt{cite} & 2,110 & 3,668 & 7 & 207 & 2.9 & 0.9 & 14.3  \\
% \texttt{cora} & \texttt{cora} &2,485 & 5,069 & 8 & 273 & 3.7 & 0.9 & 11.8  \\
% \texttt{senate} &\texttt{sen} & 1,884 & 16,662 & 110 & 82 & 12.1 & 0.5 & 4.4  \\
% \texttt{usps-3nn} & \texttt{us3} &9,298 & 21,256 & 10 & 918 & 4.4 & 1.0 & 16.8  \\
% \texttt{usps-10nn} & \texttt{us10} &9,298 & 68,381 & 10 & 925 & 13.7 & 0.9 & 9.5 \\
% \texttt{amazon} &\texttt{amaz} & 334,863 & 925,872 & 2,110 & 25 & 6.3 & 1.0 & 3.9 \\
% \texttt{dblp} &\texttt{dblp} & 317,080 & 1,049,866 & 1684 & 53 & 6.0 & 0.6 & 3.5 \\
% \texttt{friendster} & \texttt{fri} &65,608,366 & 1,806,067,135 & 3,704 & 61 & 21.0 & 0.1 & 2.9 \\
% \texttt{livejournal} & \texttt{lj} &3,997,962 & 34,681,189 & 3,442 & 38 & 22.7 & 0.6 & 2.5 \\
% \texttt{orkut} & \texttt{ork} &3,072,441 & 117,185,083 & 4,571 & 236 & 29.9 & 0.3 & 4.7 \\
% \texttt{youtube} & \texttt{yout} &1,134,890 & 2,987,624 & 1,266 & 47 & 3.4 & 0.1 & 4.4 \\
\bottomrule
\end{tabularx}
\end{centering}
\caption{\label{tab:dataset-chars}
Summary of the ground truth community properties. From left to right, the columns indicate: dataset name, abbreviated name used in later figures, number of nodes, number of edges, number of communities analyzed (we restricted analysis to communities with 10 or more nodes), average community size, average within-community degree of community members, average fraction of community member's edges that stay within the community,
and average diameter of communities.
}
\end{table*}

We summarize the dataset properties in Table~\ref{tab:dataset-chars}. We include the ground truth community datasets from the SNAP collection, the co-purchasing networks \texttt{amazon} and co-authorship graph \texttt{dblp}, as well as the social networks \texttt{friendster}, \texttt{livejournal}, \texttt{orkut}, and \texttt{youtube}~\cite{yang2012definingevaluating,mislove-2007-socialnetworks}.
These datasets are widely used as ground truth for evaluating community detection algorithms as they have thousands of annotated ground truth communities; we use the top 5,000 ground truth communities supplied by the SNAP collection for each dataset.

Additionally, we use \texttt{citeseer} and \texttt{cora}, citation networks with categories as communities~\cite{sen:aimag08}, and \texttt{senate}, a 3-NN (nearest-neighbor) network of all US senators in the first 110 congresses, where edges connect senators with similar voting patterns and communities are taken to be individual congresses; data made available by~\cite{PhysRevE.91.012821} and processed by~\cite{Kloster-2015-ppr-paths}.
Finally, we include two versions of the widely used USPS hand-written digits dataset, where nodes represent images of hand-written digits, edges are determined via a $k$-NN construction, and communities are determined by the actual digit each node represents. We use a 3-NN version, \texttt{usps-3nn}, as well as a 10-NN, \texttt{usps-10nn}, to study how such a parameter choice affects performance of the local graph analysis techniques we consider. These graphs are made available by
\cite{zhou2003learning}, though we use versions processed by~\cite{Kloster-2015-ppr-paths}.

\vpara{Processing the datasets.}
We emphasize that many of the ground truth communities have very few nodes, in the SNAP collection in particular. Furthermore, some of the $k$-NN network communities are disconnected (if we look at the subgraph induced by the set of community members), even though the full networks are themselves connected. In light of this, to make the community characteristics more meaningful, we preprocess all datasets as follows.
All datasets are first made undirected and unweighted, then the largest connected component (LCC) is extracted.
For each dataset, we restrict each community to just its members that lie inside this LCC. Any community that is not connected as an induced subgraph we separate into its connected components, and each such connected component we count as a separate community. Finally, we consider only communities that have at least 10 nodes. %The network properties in Table~\ref{tab:dataset-chars} were computed after this preprocessing.

\vpara{Computing details.}
All experiments were performed using a Dual CPU system with Intel Xeon E5-2670 processors (2.6 GHz, 8 cores) with 16 cores total and 256 GB of RAM.
All algorithms were implemented in Matlab and Matlab's C++ MEX-interface.
For timing purposes, all algorithms were run in serial, with the exception of the MOV algorithm (without subgraph extraction), which used 12 cores in parallel when performing Matlab backslash solves.

%\section{Adaptive Subgraph Extraction}

%%%
%%% SUBGRAPH EXTRACTION
%%%

\begin{figure}[htbp]
\centering
\includegraphics[width=\linewidth]{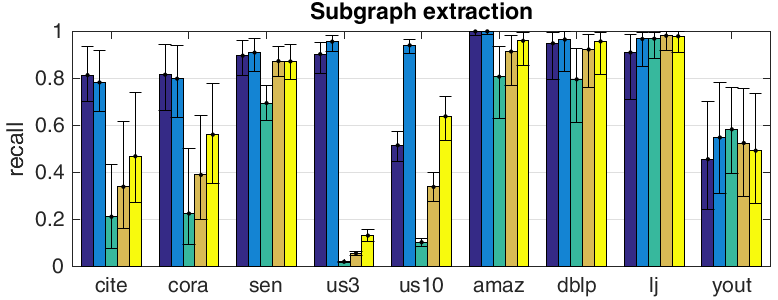}\\
\includegraphics[width=\linewidth]{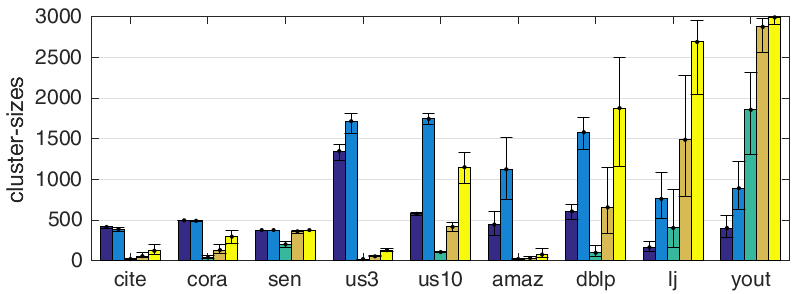}\\
\includegraphics[width=\linewidth]{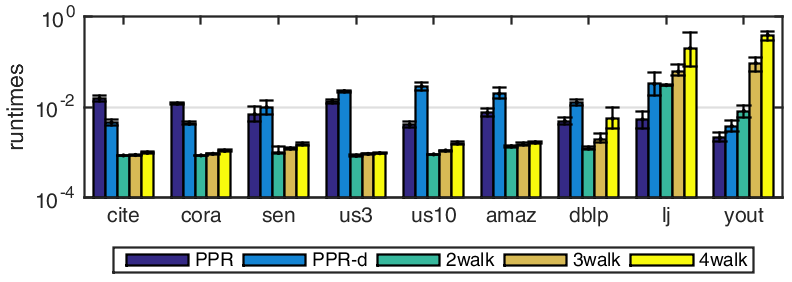}
\caption{ (\emph{Top}) Each column gives the average recall attained by various pre-processing techniques intended for extracting a good recall set of nodes near a seed node.
%The average is computed over all communities for each dataset.
 %The personalized PageRank diffusion with fixed parameters $\alpha = 0.99$ and $\eps = 10^{-4}$ outperforms the walk-based methods on datasets with larger diameter (see Table~\ref{tab:dataset-chars}). In contrast, on denser graphs with smaller diameter the walk-based methods yield better performance because the density causes PPR with fixed parameters to converge before reaching a large nodeset (visible in the \emph{middle} plot, which displays the average size of the clusters returned by the different methods). To combat this, PPR-d adapts the parameters $\alpha$ and $\eps$ so that the diffusion reaches a nodset of the desired size.
 (\emph{Middle}) The average size of the cluster returned by each method. %The plots reveal that the walk methods fail too diffuse to a good recall subgraph on the networks with larger diameter communities. In contrast, the adaptive PPR-d almost always spreads to a large enough chunk to obtain good recall.
 (\emph{Bottom}) The adaptive PageRank parameters cause a slight increase in runtime, but both PPR versions are much faster on larger graphs where even a length-3 walk can involve touching a huge number of nodes.
\label{fig:subgraph-extraction-performance}
}
\vspace{-1em}
\end{figure}

\section{Subgraph extraction with good recall}\label{sec:extraction}
%We extensively evaluate various diffusion methods including short-step random walk (used in~\cite{li2015overlapping}) and consider alternative subgraph extraction methods using personalized PageRank and Heat Kernel diffusion. We also relate the algorithms' performances to observed graph properties like edge density and community diameter.

%We first evaluate the performance of the subgraph extraction method of~\cite{li2015localspectral}, propose an altered subgraph extraction method using personalized PageRank, and relate the algorithms' performances to observed graph properties like edge density and community diameter.
In this section, we study the task of identifying a large set $T$ of nodes with {\em high recall}, from a single seed node. Given a set of ground truth nodes $C$, the recall of a proposed cluster $T$ is defined by $\text{recall}(T) =  |C\cap T| /|C|$, i.e. this is the ``{\em fraction of the truth that we obtain}". We use recall to measure how well subgraph extraction methods perform in capturing the full ground truth community.

%Formally, we address the problem of identifying a small set of ground truth nodes $T$ with {\em high precision}, from a single seed node.  Given a set of ground truth nodes, precision is defined by precision$(T) = |C\cap T|/|T|$, i.e. this is the ``{\em fraction of our guesses that are correct}".

\vpara {Baselines.} We carry out the experiment as follows. Following the subgraph extraction method of Li et al. \cite{li2015localspectral}, we perform a $k$-step walk from the seed node by computing $\bar{\mA}^k\ve_S$, where $S$ is a seed node from a given community $C$. We then take the largest 3,000 entries in the vector, and extract the subgraph corresponding to those nodes. We call the set of nodes in this subgraph $T$, and record the recall of $T$ with respect to the ground truth community $C$. For comparison, in addition to using $k = 3$ as in \cite{li2015localspectral}, we also perform the $k$-walk subgraph extraction for values $k=2$ and 4.

We also consider the personalized PageRank diffusion. Specifically, we use the ``push" implementation of PPR in~\cite{andersen2006-local} to carry out the same procedure as for the $k$-walk methods: first compute the diffusion vector starting from the seed node, then extract a subgraph using the top ranked 3,000 nodes from the diffusion vector. For PPR we used $\alpha = 0.99$, a common setting in community detection, and an accuracy of $\eps = 10^{-4}$.

It is common to normalize diffusion vectors by dividing each node's diffusion score by the node's degree. We performed this normalization on all diffusion vectors in this experiment.

The results presented are computed as follows. For each dataset, for each community, run each algorithm seeded on each individual community member as seed and average those results for that community. Then, average these community scores over all communities to obtain the results for that dataset. The errorbars indicate the standard upper and lower semi-deviations.

\vpara {Adaptive PPR subgraph extraction.} In addition to PPR, which uses fixed parameter values throughout, we propose a new variant called Adaptive PPR (or ``PPR-d''), which chooses $\eps$ and $\alpha$ so that the expected size of the output equals the target subgraph size 3,000 (or $n/5$ if $n < 3000$).
This can be accomplished by setting the desired edge-volume to equal ( desired number of nodes $)\times($ estimated average degree of community ).
For PPR this amounts to setting $1/(\eps(1-\alpha) )$ equal to the desired edge volume.
This is because the PageRank-based algorithm produces a cluster with edge volume roughly equal to $O(1/( \eps( 1-\alpha) ) )$~\cite{andersen2006-local}.
The advantage of PPR-d over PPR is that PPR-d is able to carve out larger chunks in graphs regardless of the density of the graph. For all these methods, if the subgraph reached by the walk (or diffusion) is smaller than 3,000 nodes, then simply use the full set of nodes as the extracted subgraph.

The results in Figure~\ref{fig:subgraph-extraction-performance} show that PPR-d subgraph extraction attains the best recall on almost all graphs. We notice there is an interesting trade-off in performance between PPR-d and the $k$-walk methods as the diameter and edge-density shift. The citation networks \texttt{citeseer} and \texttt{cora}, as well as the $k$-NN USPS networks, all of which have communities with large diameter, show poor performance with the $k$-walk methods. This is explained by the fact that, because of the large community diameter, a short walk is unable to reach much of the community. In contrast, on the social networks \texttt{friendster}, \texttt{youtube}, \texttt{orkut}, and \texttt{livejournal}, all of which have communities with much smaller diameter, the $k$-walk methods show much better recall, even beating PageRank in some cases.
%We find that the walk-based methods for subgraph extraction perform much better on the social networks than on the collaboration and $k$-NN graphs -- we explain this by noting that the social networks have very small diameter, enabling a small walk to stay within the community, whereas the other networks have larger diameter and so a short walk does not reach a large chunk of each community.
%While it is widely known that social networks have small diameter~\cite{leskovec2005graphs}, and
%other works have noted that certain clustering methods tend to produce compact, small diameter clusters~\cite{Leskovec-2009-community-structure}.%, to the best of our knowledge this is the first observation that ground truth communities themselves tend to have small diameter and radius, an idea we think of as the \emph{small-small world property}, in reference to the ``small world property".

Next, we note that PPR-d achieves very high recall on most datasets, in part owing to the large clusters it identifies (visible in Figure~\ref{fig:subgraph-extraction-performance}, middle plot). The datasets where PPR-d performs the worst are \texttt{orkut} and \texttt{friendster} -- this poor performance is likely due to the smaller size of clusters that PPR-d is identifying on those datasets. This smaller PageRank-cluster size is caused by the larger edge-density in those datasets: this is because for a fixed accuracy parameter $\eps$, the PPR produces a cluster with edge volume roughly equal to $O(1/( \eps( 1-\alpha) ) )$, as noted above.
From Table~\ref{tab:dataset-chars} we can see that \texttt{friendster} and \texttt{orkut} have some of the most edge-dense communities of all datasets. Interestingly, \texttt{livejournal}, which has very edge-dense communities,  does not exhibit this behavior, and PPR-d is able to attain large clusters, and good recall, here. This could be because of the larger within-community edge-ratio of \texttt{livejournal}, compared to the other dense datasets.

\vpara{Conclusion.}
We conclude that, overall, the PPR-d subgraph extraction technique attains the best recall, especially on networks with larger diameter where the $k$-walk approach used in \cite{li2015localspectral} fails to expand to much of the large-diameter community.
Furthermore, the results suggest that, regardless of which subgraph extraction technique is used, the parameters (length of $k$-walks, or accuracy $\eps$ in diffusions) should be tuned to the specific properties of the communities being sought, i.e. diameter and edge density, when possible. We discuss this in more detail in Section~\ref{sec:char}.

\vpara{Theoretical analysis.} Here we argue from a theoretical perspective why we should expecet subgraph extraction to improve the performance of local graph diffusions in community detection. Provided with a seed node $s$, suppose the subgraph extraction algorithm obtains a subgraph $G_s = (V_s, E_s)$ with high recall $(1-\epsilon)$ so that we have $|C| = (1-\epsilon) |V_s\cap C|$. Consider a $k$-step random walk on the subgraph $G_s$, starting from $s$. At step 1, the amount of probability leaving $C$ can be bounded by
\begin{equation}
\begin{split}
(\mP^1\vp_S)^T \ve_{\bar C}  & \le\text{cut}(C,\bar C) \cdot \frac{1}{\text{cut}(C,V_s)} \\
& = \text{cut}(C, V_s \backslash C) \cdot \frac{1}{\text{cut}(C,V_s)} \\
& \approx \text{cut}(C, V_s \backslash C) \cdot \frac{1}{\text{cut}(C,V)}
\end{split}
\end{equation}
If we assume subgraph size $|V_s| \ll |V|$ then the size ratio is $\Delta = |V_s| / |V| \ll 1$. The number of possible edges connecting to nodes outside the community $C$ has been approximately reduced by a factor of $\Delta^2$. Therefore,
\begin{equation}
\begin{split}
  (\mP^1\vp_S)^T \ve_{\bar C} & \le \text{cut}(C, V_s \backslash C) \cdot \frac{1}{\text{cut}(C,V)}  \\
& \approx \Delta^2 \text{cut}(C, V \backslash C) \cdot \frac{1}{\text{cut}(C,V)}  \\
& = \Delta^2 \phi(C) \ll \phi(C), \\
\end{split}
\end{equation}
where $\phi(C)$ is the escaping probability bounded on the original graph $G$.

%%%
%%%  PRECISION
%%%
\section{Seed set augmentation with good precision}\label{sec:precision}
For many semi-supervised learning tasks it can suffice to produce just a few new ground truth nodes with high precision, rather than an entire community. Given the difficulty of identifying a whole community with high quality, in this section we explore whether the algorithms we consider in this paper are reliable for the task of {\em seed set augmentation}.

Formally, we address the problem of identifying a small set $T$ of ground truth nodes with {\em high precision}, given a single seed node.  Given a set of ground truth nodes $C$, precision of the set $T$ is defined by precision$(T) = |C\cap T|/|T|$, i.e. this is the ``{\em fraction of our guesses that are correct}".
Different applications might use different values of $\tau := |T|$; we fix $\tau = 3$.

\begin{figure}[t]
\includegraphics[width=\linewidth]{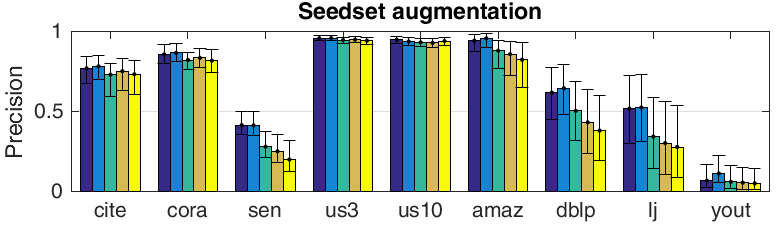}\\
\includegraphics[width=\linewidth]{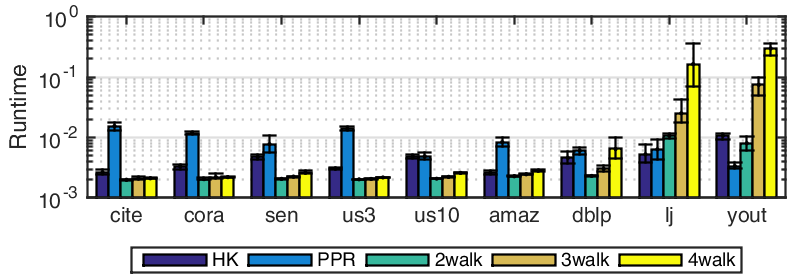}%
\caption{ {\em (Top)} Precision of trying to identify $\tau=3$ new ground truth nodes by using the ``best" $\tau$ nodes as determined by a local diffusion or walk operation. PPR is personalized PageRank with $\alpha = 0.99$ and $\eps = 10^{-4}$ whereas HK is heat kernel with $t = 4$ and $\eps = 10^{-4}$. The other methods are the random-walk vectors $\Abar^p \ve_s$ for $p=2,3,4$. {\em (Bottom)} Average running time of different methods in log scale.
}
% \vspace*{-\baselineskip}
 \label{fig:seed-precision}
\end{figure}

Figure~\ref{fig:seed-precision} (top) displays the precision of the PageRank and heat kernel diffusions alongside the same $k$-step random walks as above. For all methods, the given diffusion vector is computed from a single seed, each node's diffusion value is then divided by the node's degree, the $\tau = 3$ nodes with largest normalized score are then taken from the vector to form $T$, and finally the precision of the set $T$ is computed.

For each algorithm and each dataset, we compute the precision obtained using each community member as the seed and average those results to obtain the score for that community; then we average the scores for each community to produce the score for the dataset. The error bars give the upper and lower semi-deviations taken over the community scores. The runtimes displayed in Figure~\ref{fig:seed-precision} (bottom) were computed the same way (note the log scale).

Figure~\ref{fig:seed-precision} (top) shows that the diffusion methods offer noticeably superior precision in locating new ground truth nodes; but all of the methods fail on the datasets with worst (lowest) within-community edge-ratio\footnote{It is defined by the average fraction of edges of a community member that stay within the community.}, i.e. \texttt{lj}, \texttt{senate}, and \texttt{youtube}\footnote{Because the walk methods ran prohibitively slowly on the densest datasets, we do not include results here for \texttt{friendster} and \texttt{orkut}.}.

We later see that these same datasets (i.e. with low within-community edge-ratio) have the lowest F1-score for all algorithms. This precision experiment gives another window into why these datasets are so problematic: even the nodes that the diffusions rank as most relevant to the seed are wrong quite often when communities have low within-community edge-ratio.

\section{Our simpler LEMON}\label{sec:algorithm}

In addition to the heat kernel and PageRank diffusions and the MOV and \lem vectors that we discussed above, we introduce here our own novel improvements to the \lem method (Algorithm~\ref{alg:lemoneasy}).

To better understand the relationship of \lem~to diffusions we ran an experiment to determine which walk vector $\Abar^p \vp_0$ the \lem~procedure most heavily weighted when solving the minimization problem in~\eqref{eqn:lemon-opt}. More concretely, if $\vx = \mV_{k,l} \vy$ is the iterative \lem~vector, which entry of $\vy$ is largest in magnitude -- that is, which walk vector $\Abar^p \vp_0$ has the largest coefficient $y_j$?

To attempt to address this question, we randomly selected 100 nodes in each small dataset, and 1,000 nodes in each of \texttt{dblp}, \texttt{lj}, and \texttt{youtube}, and computed \lem vectors as follows. For each randomly selected node, we generated a set of 10 seeds using the largest ranked nodes from a local heat kernel diffusion; then we computed the \lem vector using that seed set and recording the coefficient $\vy$ such that $\vx = \mV_{3,3} \vy$.
We found that the vector $\vy$ computed in each instance placed at least $90\%$ of the weight on the walk vector $\Abar^3 \ve_S$ in over $95\%$ of trials, for all but three datasets. For the remaining three datasets (\texttt{cora}, \texttt{senate}, and \texttt{youtube}) this occurred in over $88\%$ of the trials.
% NAME    mean(y(1))     mean(y(2))      mean(y(3))    % trials with > 95% on y(1)
% cite    0.8673          0.1122          -0.0000         0.88000
% cora    0.9770          0.0302          0.0000          0.96000
% sen     0.9927          -0.0877         0.0579          0.90000
% us10    1.0000          0.0000          0.0000          1.00000
% us3     0.9904          0.0195          0.0000          0.98000
% dblp    0.9967          0.0029          0.0002          0.99600
% lj      0.9735          0.0323          0.0010          0.95600
% yout    0.8807          0.1182          0.0000          0.87300

With this in mind, we designed a modification of \lem~to be faster by avoiding the expensive optimization problem in the original algorithm that computes $\vx = \mV_{k,l} \vy$. In contrast with the original \lem, our method, which we call LEMONeasy, uses our adaptive subgraph extraction technique, and always uses $\Abar^3\ve_S$ as the iterative set-augmentation vector. Finally, we simplify the \lem stop criterion by removing their condductance-related auto-termination criterion, and instead simply perform 10 iterations of seed-set-augmentation, with a final sweep to determine the output cluster. The sweep uses an ordering on the nodes determined by the order in which nodes are added to the seed set, rather than simply by the values of the final iterative \textsc{Lemon} vector.

This framework is much simpler to implement than the original \lem~(since it no longer relies on solving a convex optimization problem), but uses the same key ideas of subgraph extraction and iterative set augmentation. Our experiments in Section~\ref{sec:precision} confirm that simple diffusion vectors are reasonably good at identifying a small number of ground truth nodes with good precision; our LEMONeasy leverages this idea to iteratively build up a set of nodes with high precision.

\floatname{algorithm}{Algorithm}\label{alg:lemoneasy}
\renewcommand{\algorithmicrequire}{\textbf{Input:}}
\renewcommand{\algorithmicensure}{\textbf{Output:}}
\begin{algorithm}[htbp]
  \caption{LEMONeasy($G,s,r,f$)}
  \begin{algorithmic}[1]
    \Require{graph $G$, seed node $s$, augment steps $r$, augment size $f$}
    \State Initialize $\vp_0$.
    \State Extract subgraph $G_s = (V_s, E_s)$ starting from $s$, using method in Section~\ref{sec:extraction}.
    \State Initialize stack $S_0 = [s] $
    \State For $j=0$ to $r$, push onto $S_j$ the top $j\cdot f$ nodes of the vector $\vz = \Abar^3 \ve_{S_j}$, largest $z_i$ first.
    \State Perform sweep over the nodes in the stack $S_r$ plus the remaining nodes of $G_s$ ordered by $\Abar^3 \ve{S_r}.$
    \Ensure {The best conductance set found by the sweep.}
  \end{algorithmic}
\end{algorithm}

%%%
%%%  GROUND TRUTH RECOVERY
%%%
\section{Recovering ground truth communities} \label{sec:experiments}
Here we evaluate the algorithms' ability to identify a ground truth community given a single seed from that community.
All performance results reported in this section are computed as follows. For each dataset, for each community, use each community member as an individual seed and run each algorithm from it; then average the performance over each seed to obtain a score for the community. Finally, compute the average results over all communities to obtain the score displayed for each dataset. In this manner, the results we report here reflect the expected performance of each algorithm if a seed node were chosen uniformly at random from a community of nontrivial size that was chosen uniformly at random. Error bars in the plots indicate the standard upper and lower semi-deviations from the average.

\vpara{Effect of subgraph extraction.}

\begin{figure}[t]
  \includegraphics[width=\linewidth]{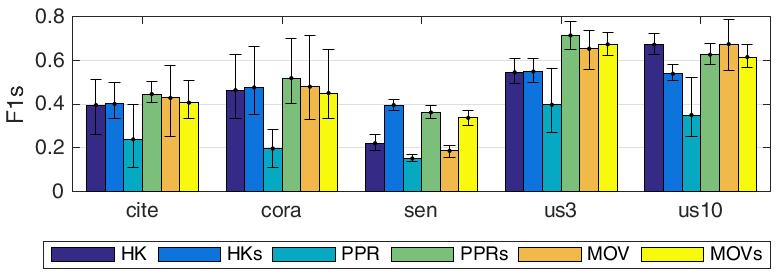}\\
  \includegraphics[width=\linewidth]{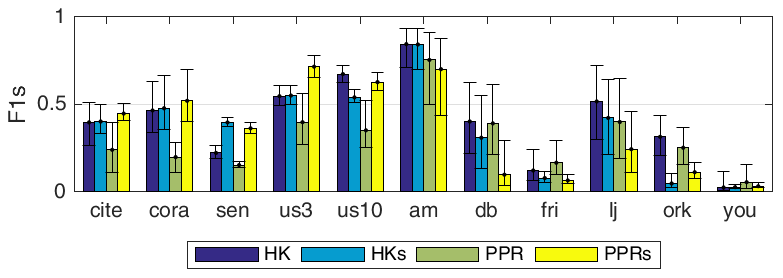}%
  \caption{ Effect of subgraph extraction on the F1 score attained by baseline codes (HK, PPR, MOV). (An ``s" in the name indicates subgraph-extraction is used.) ({\em Top.}) Comparison of MOV and MOVs on the smaller datasets shows a general decrease in the deviation of F1 scores, an increase in performance for \texttt{senate}, and little other change. The performance of PPRs generally increases, in some cases substantially, but for HKs there is no consistent affect on performance. ({\em Bottom}.) In contrast with the small datasets, on the larger datasets the subgraph extraction procedure generally decreases performance, in some cases substantially.
  }
  \vspace*{-\baselineskip}
  \label{fig:gtc-F1-baseline}
\end{figure}

To understand the effect that our subgraph extraction technique has on the different algorithms, we display in Figure~\ref{fig:gtc-F1-baseline} the performance of HK, PPR, and MOV both with and without subgraph extraction. The top plot shows that, on the smaller datasets, subgraph extraction noticeably improves the cluster quality of PPR without substantially affecting HK and MOV in most cases. We emphasize that ``MOV" indicates the algorithm ran on 12 processors in parallel, whereas ``MOVs'' operated in serial.

\begin{figure}[t]
\includegraphics[width=\linewidth]{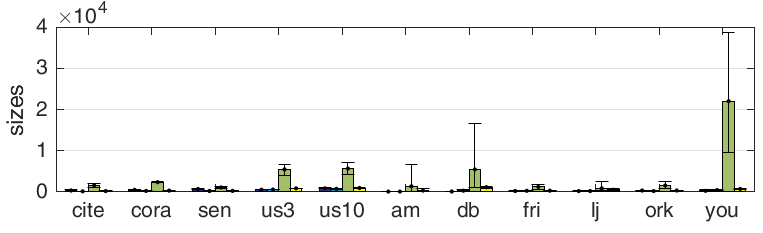}\\
\includegraphics[width=\linewidth]{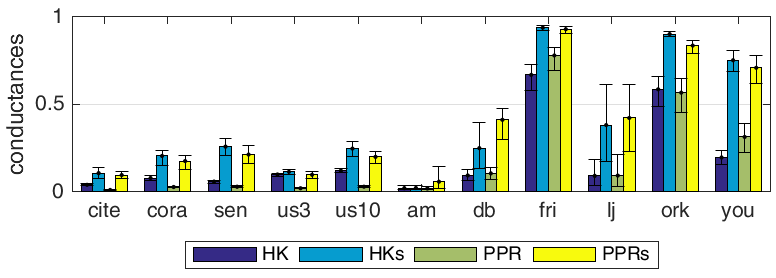}%
\caption{ Subgraph extraction decreases the size (top plot) and increases the conductance (bottom plot) of the clusters output by the baseline codes (HK, PPR), often substantially. (An ``s" in the name indicates subgraph-extraction is used.) This suggests that the difference in F1 performance of HK and PPR depicted in Figure~\ref{fig:gtc-F1-baseline} is caused by the subgraph extraction technique cutting out so much of the whole graph that the underlying good conductance cut is no longer good conducatnce, and the corresponding cluster is obscured.
}
 \vspace*{-\baselineskip}
 \label{fig:gtc-subgraph-explain}
\end{figure}

In contrast, Figure~\ref{fig:gtc-F1-baseline} (bottom) shows that HK and PPR without subgraph extraction performs better than HKs and PPRs. We did not compute MOV on the larger datasets without using subgraph extraction for comparison because, as a global method, it was prohibitively slow.
To explain this discrepancy (improvements on small graphs, but not on large), we look at the size and conductance of the clusters returned before and after subgraph extraction, in Figure~\ref{fig:gtc-subgraph-explain}.

A plausible explanation is that a set of good conductance can be missed if that set has worse conductance when computed with respect to the extracted subgraph. Without subgraph extraction, PageRank was clumping together two or more clusters into one, even better conductance cluster (as suggested by both the conductance and size plots); but with subgraph extraction, that large set no longer had good conductance, and PageRank could instead identify the true community better using conductance.
To combat this effect, during the sweep step of the algorithm, the conductance could be computed with respect to the full graph, instead of the subgraph.

\begin{figure}[t]
  \includegraphics[width=\linewidth]{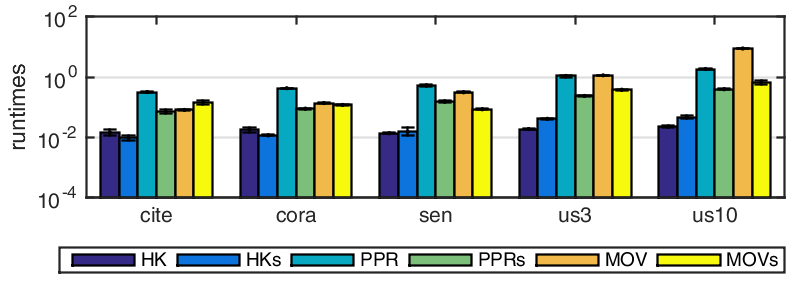}\\
\caption{ Our subgraph extraction method enables more sophisticated methods like MOV to run on larger datasets, with competitive quality.
}
 \vspace*{-\baselineskip}
 \label{fig:gtc-runtime-baseline}
\end{figure}

We remark that subgraph extraction is primarily intended to enable the use of more sophisticated, global algorithms like MOV and our method LEMONeasy. Figure~\ref{fig:gtc-runtime-baseline} shows that MOV, even running on 12 processors in parallel, is still slower on the medium-sized datasets than MOVs running in serial.%; on larger datasets (bottom plot), MOVs is competitive with the local methods.

Because of how slow MOV runs on large datasets, we did not carry out thorough experiments to demonstrate the speed-up yielded by our MOVs on the largest sets. But, to give an idea, we ran our algorithm for MOV on one randomly selected community (13 nodes) in \texttt{dblp}; the average runtime to compute MOV seeded on a single member of the community was over 4,100 seconds, using 12 processors in parallel.
In contrast, MOVs averaged 0.78 seconds across all communities (running on a single processor).
(Because MOV is a global algorithm, this runtime should not vary much across different seeds or different communities.)

\begin{figure}[t]
\includegraphics[width=\linewidth]{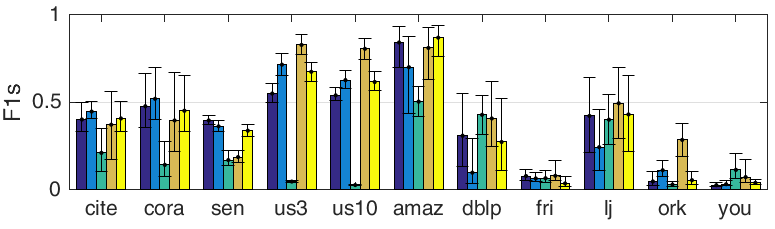}\\
  \includegraphics[width=\linewidth]{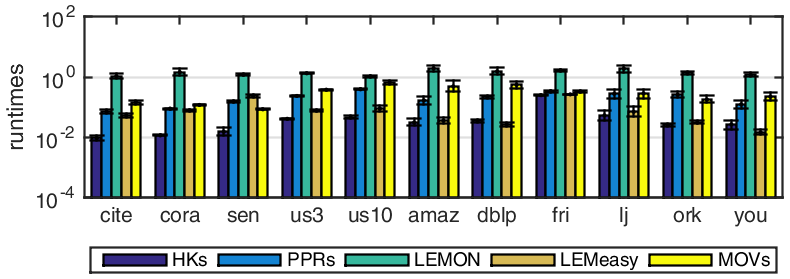}%
  \caption{ Comparison of new techniques (LEMONeasy, and MOV with our subgraph extraction) with baseline codes. Our subgraph extraction enables the first such evaluation of MOV on large datasets. Our modifications to \lem result in a two-orders-of-magnitude speed-up on large datasets, with competitive cluster quality. We remark that HKs and PPRs perform poorly on larger datasets; instead, consult Figure~\ref{fig:gtc-F1-baseline} (bottom) for the performance of HK and PPR (without subgraph extraction) on large datasets.
  }
   \vspace*{-\baselineskip}
   \label{fig:gtc-F1-performance}
\end{figure}

\vpara{Ground truth recovery peformance}
Finally, Figure~\ref{fig:gtc-F1-performance} compares the performance of our novel modified versions of previous algorithms (LEMONeasy and MOVs), with baseline approaches. Our LEMONeasy attains a two orders of magnitude speed up compared to the original proposal by Li et al.~\cite{li2015localspectral}, and obtains cluster quality competitive with all baseline codes. The running time of our improved version MOVs is competitive even with the local methods on larger datasets (bottom plot).

%show the effectiveness of using subgraph extraction technique to enable MOV on large datasets. Using our pre-processing technique speeds up MOV by an order of two magnitudes on large datasets, resulting in competitive cluster quality.

\section{Discussion}\label{sec:char}
%\vpara{Notable Characteristics.}
\vpara{Characteristics of datasets.} First we remark on variety of characteristics across the 11 datasets: the graph size and density and the average community diameter, size, and ``average fraction of community member's edges that stay inside the community".
%In later sections we refer back to these statistics in Table~\ref{tab:dataset-chars} to explain trends in our experimental evaluation of subgraph extraction, seed set augmentation, and finally ground truth community detection.

For example, in Section~\ref{sec:extraction} we find that the walk-based methods for subgraph extraction perform much better on the social networks than on the collaboration and $k$-NN graphs -- we explain this by noting that the social networks have very small diameter, enabling a small walk to stay within the community, whereas the other networks have larger diameter and so a short walk does not reach a large chunk of each community.

Also, we notice that the average community size ranges from 23 (Amazon) to almost 1,000 (in the USPS graphs). The study of Li et al.~\cite{li2015localspectral} showed that using loose lower and upper bounds on the average size of a community can improve the quality of clusters output by an algorithm. In the absence of such knowledge, an algorithm must be robust to such wide variation in size to be able maintain quality performance.

Next, we observe that the difference in graph construction parameters for the 3-NN and 10-NN versions of the USPS graph lead to some significant differences in community structure. The denser 10-NN version has communities with significantly lower diameter; we see that this leads to significant differences in the recall of $k$-walk based subgraph extraction on the two graphs.

%Rather than measure the average conductance of communities, instead we give the average fraction of community members' edges that stay within a community. That is, for a single community, compute the average ratio (taken over all members of the community) of a nodes' within-community edges to total edges; then for each graph, average that number over all communities. Whereas conductance is an aggregate measure of a community's internal cohesion versus external connectivity, this metric is an average of a local metric, and better measures ``how good is each node as a seed node". We see that this metric is a good predictor of algorithm failure -- that is, a low value for average within-community edge-ratio is a reliable indicator that all algorithms output clusters of poor quality.

%Finally, we note that, while it is widely known that social networks have small diameter~\cite{leskovec2005graphs}, and other works have noted that certain clustering methods tend to produce compact, small diameter clusters~\cite{Leskovec-2009-community-structure}, to the best of our knowledge this is the first observation that ground truth communities themselves tend to have small diameter (far right column in Table~\ref{tab:data-summary}). We refer to this fact later when discussing the usefulness of short-length random walks vs local diffusions in extracting high-recall subgraphs near a seed set inside a small-diameter community (Section~\ref{ref:subgraph-extract}).

%%
%% Subgraph extraction
%%

\begin{figure}[t]
\includegraphics[width=0.8\linewidth]{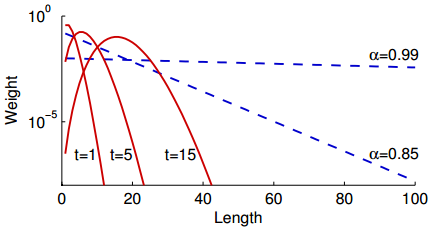}
\caption{ Diffusion coefficients ($c_k$) of personalized PageRank (dotted blue lines) and heat kernel (red). The plot shows that heat kernel places most of its weight on a small set of short-length walks, which can cause the diffusion to identify smaller clusters.
% give that of personalized PageRank with different values of $\alpha$, and the red show Heat Kernel diffusion coefficients with different values of $t$.
See Section~\ref{sec:diffusion-overview} for details of the parameters $\alpha$ and $t$. }
 \vspace*{-\baselineskip}
 \label{fig:diffusion-length}
\end{figure}

%%
%%  ALGORITHMIC CLUSTERS
%%
\vpara{Properties of algorithmic clusters.}
We also discuss here how the different diffusion algorithms that we consider respond to the dataset characteristics discussed above. We begin by discussing the traits of the clusters identified by the methods.
As an example of how different methods find clusters with different characteristics, we point to two recent studies that found that different sets of diffusion coefficients can be selected to control the size of the output clusters, to better match the size of the communities expected to be found in a given network~\cite{avron2015community,kloster2014heatkernel}. Figure~\ref{fig:diffusion-length}, reused from~\cite{kloster2014heatkernel}, shows that the heat kernel diffusion puts more weight on shorter walks, and so emphasizes smaller communities than PageRank does.

In this paper, however, we have considered not only the size but also the conductance of the output clusters, and how the different local methods are affected by the diameter and density of the underlying graph.
%In this section, we show that a subgraph extraction technique presented by Li et al.~\cite{li2015localspectral}.

%\vpara{Local output versus local algorithm.}
%* slow down from MOV or LEMON on larger graphs
%* slow down from $k$-walks on small diameter graphs
%* in contrast, PPR and HK are always consant size output.

\vpara{Community density, graph construction, and tuning diffusion parameters.}
If PageRank (or heat kernel) uses fixed parameter settings then the clusters identified will be smaller on denser datasets. This might sound counter-intuitive (i.e. ``shouldn't a denser graph spread a diffusion more quickly?"), but this fits in with the theoretical properties of the algorithms as follows. The papers that introduced both algorithms give theoretical bounds on the amount of work performed: PageRank touches no more than $O( 1/ (\eps(1-\alpha)))$ edges and heat kernel roughly $O(e^t t /\eps)$ edges \footnote{This is a simplification of the actual bound derived in~\cite{kloster2016graph}}. Thus, for fixed parameter settings, these two diffusions can touch no more than a fixed constant number of edges. On denser graphs, there will be fewer total nodes attached to a set of edges of constant size, and so clusters with fewer nodes will be output.

This fact can have an interesting consequence when a graph is constructed from a single dataset in different ways. As a case study, we consider two graphs constructed from the same USPS hand-written digits dataset. We find that MOV and PageRank spread to larger portions of the graph, and so consider a larger number of possible sweep-cuts --- thus, MOV and PageRank find the lowest conductance sets out of any algorithm, but also potentially the largest sets. At the same time, the tendency of these algorithms to explore such large chunks of the graph means they tend to require more time than the $k$-step walk or heat kernel methods.

On the other hand, $k$-step random walks, PageRank and heat kernel diffusion vectors have non-zero weights on paths of potentially excessive length ($k>15)$, thereby diffusing probabilities to a larger chunk of the graph and encouraging large communities. In contrast, using $k$-step random walks can miss out on potentially important portions of the graph if $k$ is not large enough (e.g., small relative to the average community diameter).

\section{Conclusions}\label{sec:conclusions}
%This paragraph will end the body of this sample document.
%Remember that you might still have Acknowledgments or
%Appendices; brief samples of these
%follow.  There is still the Bibliography to deal with; and
%we will make a disclaimer about that here: with the exception
%of the reference to the \LaTeX\ book, the citations in
%this paper are to articles which have nothing to
%do with the present subject and are used as
%examples only.

Our results suggest that community and graph characteristics must be taken into account when selecting algorithms for semi-supervised learning tasks, as well as parameters. In particular, the details of a graph's construction (e.g. 3-nearest-neighbor vs 10-NN), when known, should lead to differences in algorithm choice and parameter selection.
 Our experiments with PageRank-based subgraph extraction show that adapting parameters can lead to consistent performance across varied networks, and that subgraph extraction can enable rapid application of more sophisticated techniques that would otherwise be intractable.

%\TODO The problem of growing a single seed into a set of relevant nodes is a difficult problem that has been a thriving area of research of late.
% We contribute to this literature by proposing and evaluating techniques for identifying both medium-sized high recall sets and very small, high precision sets of nodes. Our techniques are highly useful in conjunction with other local graph analysis algorithms as fast pre-processing routines, but are also of interest in their own right. Our novel method for seed set augmentation will be useful in any application that benefits from having a larger seed set.

% Our analysis of community properties, namely edge density, diameter, and radius, partially explain the difference in performance between the PageRank and walk based subgraph extraction methods that we analyze. In particular, the walk-based approach fails in networks where communities have larger diameter. However we observe that communities in social networks tend to have very small diameter, and so walk-based approaches perform much better; at the same time, these social networks have larger edge density than $k$-NN networks, and so algorithms, like the commonly used PageRank ``Push" algorithm, that terminate based on edge-volume must alter their parameters according to the edge densities of the dataset in question.
We also observed that the recent, successful \lem method for local community detection, despite some algorithmic sophistication, effectively uses a simple walk vector to expand a seed set. We used this insight to present a greatly simplified algorithm that iteratively grows a seed set with a simple walk vector. Our experiments on the precision of different methods found that diffusions outperformed walk vectors; in the future, we plan to explore the effect of using such diffusions in the \lem framework instead of the lower precision walk vector.

\section*{Acknowledgments}
This work was supported by Purdue University, NSF CAREER award CCF-1149756, and US Army Research Office %W911NF-14-1-0477. The first author completed much of his contribution to this paper while at Purdue University.

\bibliographystyle{abbrv}
\bibliography{main}

\end{document}